


\font\twelvebf=cmbx10 scaled 1400

\raggedbottom

\def\rref#1/{$^{#1}$}


\newcount\refno
\refno=0
\gdef\refiss{\global\advance\refno by1\item{\the\refno.\ }}

\def\totalno{\message
  {***Total:(\the\refno) references***}}


\input jnl
\input epsf.tex

\rightline{LA-UR-92-3847}
\centerline{\twelvebf Time Symmetric Quantum Cosmology}
\centerline{\twelvebf and }
\centerline{\twelvebf Our Universe}
\vskip 1 truein

\author Raymond Laflamme\footnote{$^*$}{Email: %
laf@tdo-serv.lanl.gov.}

\affil
Theoretical Astrophysics, T-6, MSB288
Los Alamos National Laboratory
Los Alamos, NM 87545, USA
\vskip 0.2in
\centerline {December 1992}
\abstract
We investigate the time neutral formulation of quantum cosmology of Gell-Mann
and Hartle. In particular we study the proposal discussed by them that our
Universe
corresponds to the time symmetric decoherence functional with initial and final
density matrix of low entropy.   We show that our Universe does not correspond
to this proposal by investigating the behaviour of small
inhomogeneous perturbations around a Friedman-Robertson-Walker model. These
perturbations cannot be time symmetric if they were small at the Big Bang.
\endtopmatter
\endpage

The  origin of the arrow of time is one of the fundamental problem of physics.
The world around us has a definite arrow of time.  We have often seen cups
falling off tables and breaking into a multitude of pieces but the time
reversed
situation as seen from a movie played back in time certainly does
corresponds to our everyday experience.  This is a puzzle as the laws of
physics we know are CPT invariant, and thus time invariant for  CP invariant
matter.  So what is the origin of the arrow of time?

In order to study this question it is important to distinguish different arrows
of time and see how they can be related.
First there is the  thermodynamical arrow of time
encapsulated in the second law of thermodynamics.
It asserts that the entropy of a system cannot decrease.
Secondly there is a cosmological arrow of time; the fact that  the
Universe is expanding.  This is deduced from the observation that the  spectrum
of distant galaxies are redshifted.  In the early sixties, Gold\rref1/
suggested
that these two arrows had to be correlated and postulated that the
thermodynamical  arrow of time would even reverse during the recontraction of
the Universe.   This intriguing possibility seems to have resurfaced many times
in physics.  Other arrows of time are also implied by the
use of retarded potential in electromagnetic theory instead of advanced ones
and the psychological one, the feeling that time flows in one direction.
However these arrows are believed to follow from the thermodynamical one, so we
will not mention them anymore in this letter.

All these arrows can be derived from specific initial conditions of the
Universe\rref2/.   However there is also a different arrow which appears in
quantum mechanics  through the process of measurements\rref3,4/. This arrow is
related to
the collapse of the wave function. This arrow seems to have a different status
to the other one mentioned above as its origin is in the dynamical law.
Aharonov, Bergmann generalised quantum mechanics so that it does not have this
inherent time
asymmetry by introducing both initial and final conditions in the probability
formula of Quantum Mechanics. Recently Gell-Mann and Hartle\rref6/ have
proposed a generalisation of this time neutral theory of quantum mechanics
which
includes closed system. They  are hoping that it might be instructive to
generalize Quantum Mecahnics so that it does not blatantly distinguish between
the past and the future.  In fact, in this formulation even the quantum
mechanical arrow of time
is a consequence of specific boundary conditions, in this case a particular
initial and final density matrix.
They have discussed a time neutral formulation of
quantum cosmology where both the initial and the  final density matrix have low
entropy. This could be thought of as a mathematical realisation of Gold's idea.

The purpose
of this letter is to investigate what are the observable consequences of this
completely symmetric boundary condition and to show that our Universe does not
correspond to this proposal.   A low initial entropy is incompatible with a
final low entropy for our Universe.  In order to demonstrate  this assertion we
will study
Friedmann-Robertson-Walker model with small inhomogeneous perturbations
using the Einstein-Hilbert action.  A coarse grained entropy could be obtained
by averaging the field over a given volume of space. However we
will simplify the problem by associating a low entropy Universe with one which
has small inhomogeneous perturbations and mutadis mutandis for large entropy.
This is reasonable as small perturbations have a quadratic Lagrangian
and thus constant entropy in a given volume.  If the perturbations become
large the cubic part of the Lagrangian  will couple all modes of different
wave numbers and thus the entropy in a giveWe  will investigate the behavior of
small inhomogeneities and show that if they
start with small values (when the scale factor of the Universe is small) it
is impossible for them to return to small values at the end of the
recontracting phase.  In fact they will blow up at the endpoint of
recontraction.  Thus a low entropy Universe at the Big Bang cannot come back to
low entropy at the Big Crunch.
\vfill\eject

\proclaim The time neutral formulation of quantum mechanics and cosmology.

Let's first review how an arrow of time is introduced in quantum mechanics and
how it is  possible to formulate this theory in a time neutral way.

Let $\{\alpha_k\}$ be an exhaustive set of alternatives at time $t_k$
represented by the set of projectors $\{ P^k_{\alpha_k}(t_k)\}$ in the
Heisenberg representation.  The probability for an history in the exhaustive
set is given
by
$$
p(\alpha_n, ...,\alpha_1)=
Tr \Big[ P^n_{\alpha_n}(t_n) ... P^1_{\alpha_1}(t_1) \rho_i
         P^1_{\alpha_1}(t_1) ...P^n_{\alpha_n}(t_n)  \Big]
\eqno (1)
$$
where  $\rho_i$ is the state at the initial time.  The time asymmetry of this
formulation doesn't come through the time ordered product of the operator
$P^k_{\alpha_k}(t_k)$.  To a sequence of these ordered  product there always
exists a CPT transformed one which leaves the probability invariant.
The arrow of time comes form the  fact that there is an initial
density matrix $\rho_i$  (an initial condition) but no final density matrix
(final condition).

Aharonov, Bergman and Lebovitz\rref5/ transformed this theory so that it
becomes time neutral  by adding a final density matrix in the probability for
an history
$$
p(\alpha_n, ...,\alpha_1)=
Tr \big[\rho_f P^n_{\alpha_n}(t_n) ... P^1_{\alpha_1}(t_1) \rho_i
               P^1_{\alpha_1}(t_1) ...P^n_{\alpha_n}(t_n)  \big]
    \Big/Tr[\rho_f \rho_i]
\eqno (2)
$$

Gell-Mann and Hartle have generalised the above theory so that it could be
applied
to closed system. In this case a necessary condition to assign probabilities to
histories is that the decoherence functional between two histories
${\cal C}_\alpha,{\cal C}_{\alpha'}$ characterised by different strings of
projectors vanishes.
$$
D_{\alpha,\alpha'}
    \equiv Tr[ \rho_f {\cal C}_\alpha \rho_i {\cal C}^dagger_{\alpha'}]
     \Big/Tr[\rho_f \rho_i]
     = 0
\eqno (3)
$$
If this is true for all $\alpha\neq\alpha'$, then it is possible to assign
probability which will satisfy the sum rule for each history $\alpha$.

They have investigated the limitations imposed by decoherence and classicality
on this time neutral formulation of quantum mechanics and quantum cosmology.

They first investigated the limitations due to the decoherence requirement.
They showed that if both $\rho_i$ and $\rho_f$ are  pure states,  decoherence
doesn't occur unless there are at most two vanishing quantities
$<\Psi_i | {\cal C}_\alpha |\Psi_f>$ and thus there  are at most two decoherent
coarse grained histories.  They also showed that the case $\rho_i=\rho_f \equiv
\rho$ was of marginal interest.  In this case the system decoheres only if
$$
 [ {\cal C}_\alpha,\bar\rho] = 0 \ \ {\rm for\ all\ } \alpha
\eqno (4)
$$
with $\bar\rho = (Tr\rho^2)^{-1} \rho^2$.  This implies a trivial dynamics and
therefore not very interesting.

Once a coarse graining has been selected and decoherence has occurred they
noted that there are still restrictions imposed by the classical behavior.
In particular they considered  a classical two-time
boundary problem: the Ehrenfest urn problem with initial and final conditions.
It consists of $n$ numbered balls disposed in two urns.  The dynamical law is
provided by transferring from one urn to the other the ball corresponding to a
number becounting only the number of balls in each urn. They studied the case
where all
the balls were initially and finally in only one urn.
They concluded that if the relaxation time is much shorter than the time
interval of interest it is not  possible to distinguish between a
time-symmetric
and other solutions.  They suggested therefore that if the time of maximum
expansion is very large than the present observed time asymmetry could still be
consistent with a time symmetric Universe.  A measure of how probable is a
time symmetric Universe given initial and final density matrices is
$$
N^{-1} = Tr(\rho_f\rho_i)
\eqno (5)
$$
the  fraction of trajectories meeting the initial condition
that  also meet the final one
We  show below that this is  incredibly small in the cosmological case where
initial and
final density matrix have low entropy. Moreover the trajectories which do
meet the initial and final conditions would lead to a very uninteresting
Universe.

\proclaim Cosmological model.

Let's first assume that the Universe is in a
quasi-classical domain.  Thus the time neutral quantum mechanical
formulation  reduce to a classical two-time boundary condition.  For the
purpose of this letter we study a rather simple cosmological model which
nevertheless contains the essential features of more general model.

Let's assume that  the Universe contains a perfect fluid with no pressure
(dust).   This should be a good approximation as long as we do not probe the
fundamental field which give rise to this equation of state to a scale
smaller than its Compton wavelength\rref2/.  Thus the scale factor of the
Universe behaves as
$$
a= {a_m\over 2} ( 1-\cos\eta), \ \ \ \  \eta\in\lbrack 0,2\pi \rbrack
\eqno (6)
$$
where $a_m$ is the scale factor of the Universe at maximum expansion and $\eta$
is tUniverse was radiation dominated or inflationary.  These phases will only
change the behavior of small inhomogeneities quantitatively but not
qualitatively.

There are two types of perturbations we can study; the gravitons (the
transverse and traceless part of the metric) and the scalar perturbations
coming from the fluid.  The fact that our Universe is not time symmetric is
best seen from the
scalar perturbations.

The scalar perturbations are mixtures of the scalar part of the metric and of
the field which makes the fluid.  A way to take care of the gauge degree
of freedom is to use a gauge invariant formalism.
For example the gauge invariant version of the lapse perturbation $\Phi$ (see
ref.[7]) obey the equation
$$
\Phi'' + 3 {\cal H} \Phi' + [2{\cal H}' + {\cal H} -1]\Phi =0
\eqno (7)
$$
where ${\cal H} =a'/a$.  The solution are given by
$$
\Phi(x,\eta) = C_1(x) { \sin^2\eta -3\eta\sin\eta -4\cos\eta +4 \over
                        (1-\cos\eta)^3}
              +C_2(x) {\sin\eta \over (1-\cos\eta)^3}
\eqno (8)
$$
with $\int C_i(x)\sqrt\gamma d^3x =0 \ \ i=1,2$.
At the beginning of the Big Bang (as $\eta\rightarrow 0$) $\Phi \sim cte$ or
$a^{-5/2}$. The first solution is  chosen if we asked for small perturbations
in the early Universe.
The scalar amplitude will keep increasing throughout the history of the
Universe and will
become non-linear at some time depending on the initial amplitude.  Figure 1
shows the behavior of the scalar perturbation which is regular at the Big Bang
and blow up at the Big Crunch. It is possible to get a time symmetric solution
but in order to get it we
need the inhomogeneous perturbations to blow up both at the Big Bang and
the Big Crunch. An analysis of the case where the fluid r
Using eq.(8) it is possible to estimate the probability that
a trajectory picked from a low entropy distribution given by $\rho_i$
will also meet a final low entropy distribution.  Let's assume that the
phase space distribution of $\rho_i$ and $\rho_f$ are circles of unit
radii (in natural units). Let's also assume that we cut-off the radius
of the Universe at the Planck length $\ell_p$. The fraction of trajectories
meeting the initial condition that also meet the final condition is given by
$\sim (\ell_p / a_m)^{5/2} < 10^{-157}$ (where we assumed a minimum value for
$a_m$ (the radius of the Universe just before recollapse) as the radius fo the
present observable universe).
A rather small probability.  The trajectories which do meet the final
conditions
are the one with incredibly small initial amplitude.  These perturbations will
remain linear and will not be able to form all the  structure we see in the
Universe.

A similar conclusion can be obtained in a non-linear regime by studying a
homogeneous minisuperspace model with 3-surfaces of topology $S^1\times S^2$
instead of $S^3$ for the  FRW model.  In this case the radius of the  $S^1$
will be monotonically increasing throughout the history of the Universe if it
started with small value at the Big Bang \rref8/.

\proclaim Conclusion.

It seems that our Universe was very homogeneous in the past\rref9/.  Therefore
we can conclude that the inhomogeneous perturbations were very small just after
the Big Bang.  Thus from the study  of their behavior during the whole history
of a closed Universe we have shown that they can not be time symmetric with
respect  to the time  of maximum  expansion.  We should point out that this
assumes the cosmological principle\footnote{$^*$}{We would like to thank W.G.
Unruh for pointing this to usee in fact the history of our own region.}  From
this assumption we can conclude that perturbations were small in the early
Universe because the
microwave background has very small anisotropies.  If we accept this we can
therefore conclude that our Universe cannot correspond to the decoherence
functional with both initial and final density matrix of low entropy.

\head {Acknowledgments}

The author acknowledges the support of a Peterhouse College Fellowship
at the University of Cambridge and the Department of Energy.  He would also
like to thanks the Aspen Center for Physics where part of this research was
carried out.

\endpage

\head {References}

\refiss
T. Gold, in:
{Onzi\`eme Conseil de l'Institut International de Physique Solvay},
{\it La Structure et l'Evolution de l'Univers}(Edition Stoops, Brussels. 1958),
p 81; Am. J. Phys {\bf 30}, 403 (1962).

\refiss
S.W. Hawking, R. Laflamme and G. Lyons,
``Origin of Time Asymmetry'',
DAMTP report GR-22, University of Cambridge, 1992.

\refiss
J. von Neumann,
{\it Mathematical Foundations of Quantum Mechanics}, translated by R.T.Berger
(Princeton University Press, Princeton 1955).

\refiss
D. Bohm, {\it Quantum Theory}, (Prentice-Hall Inc, Englewood Cliffs, New Jersey
1951).

\refiss
Y. Aharonov, P. Bergamm and J.L. Lebowitz,
\pr 134, B1410, 1964.

\refiss
M. Gell-Mann and J.B. Hartle,
Time Symmetry and Asymmetry in Quantum Mechanics and Quantum Cosmology,
to published in the Proceedings of the Nato Workshop on the
{\it Physical Origin of Time Asymmetry} Maz\`agon, Spain, September30-October
4,
1991  ed. by J.J. Halliwell, J.Perez-Mercadez and W. Zurek,  Cambridge
University Press, Cambridge (1992).

\refiss
Mukhanov, V.F., Feldman H.A. and Brandenberger, R.H.;
``Theory of CCITA preprint 1991.

\refiss
Kantowksi, R.  and Sachs, R.K.;
\jmp 7, 443, 1966;
R. Laflamme,
Ph.D. Thesis, Cambridge University, 1988.

\refiss
Smoot G. and al. Ap.J. {\bf 396}, L1, 1992.

\totalno

\vfill\eject

\epsfbox{pic.ps}

\endit